# Real-Time Phase Correction based on FPGA in the Beam Position and Phase Measurement System

Xingshun Gao, Lei Zhao, *Member, IEEE*, Jinxin Liu, Zouyi Jiang, Xiaofang Hu, Shubin Liu, *Member, IEEE*, Qi An, *Member, IEEE*

***Abstract*—A fully digital beam position and phase measurement (BPPM) system was designed for the linear accelerator (LINAC) in Accelerator Driven Sub-critical System (ADS) in China. Phase information is obtained from the summed signals from four pick-ups of the Beam Position Monitor (BPM). Considering that the delay variations of different analog circuit channels would introduce phase measurement errors, we propose a new method to tune the digital waveforms of four channels before summation and achieve real-time error correction. The process is based on the vector rotation method and implemented within one single Field Programmable Gate Array (FPGA) device. Tests were conducted to evaluate this correction method and the results indicate that a phase correction precision better than ± 0.3 ° over the dynamic range from -60 dBm to 0 dBm is achieved.**

## I. Introduction

The Chinese ADS aims at exploring a safe and efficient method to produce clean energy with nuclear waste. The proton LINAC is an important component of ADS [1]. Beam diagnostics is essential for the operation of this LINAC, in which precise beam phase and position measurement is an important task. Capacitive type BPMs (CBPMs) [2] are chosen to pick up the beam signals for beam measurement. We designed beam measurement electronics which integrates the functionality of beam phase and position measurement within one single instrument [3] [4]. The system we designed imports narrow pulses from four pick-ups of one BPM with a repetition rate of 162.5 MHz and a dynamic range from -44 dBm to -4 dBm as shown in Fig. 1. As shown in Fig. 2, the beam signals from the BPM pass through amplification and Band Pass Filter (BPF) circuits, and then are digitized by Analog-to-Digital Converters (ADCs) in the Analog Front End module (AFE). To guarantee a good signal-to-noise ratio in a large dynamic range (the system design goal is ~ 60 dB), the signals are amplified to fit the Full Scale Range (FSR) of the ADCs. And the circuit gain is decided by the cascaded Radio Frequency (RF) amplifiers and digitally controlled variable attenuators. The amplified signals are then directly digitized by 16-bit high-speed ADCs based on the RF signal In-phase & Quadrature-phase (IQ) under-sampling method (162.5 MHz input signal is sampled with 50 MHz clock) [3]. Then digital Intermediate Frequency (IF) signals of 12.5 MHz can be obtained, and in each period of the IF signal, there exist exactly four samples (i.e. I, Q, -I, and –Q). These digital signals are transferred to the Digital Processing Module (DPM), and the beam phase and amplitude can be calculated with these I and Q samples as in (1) and (2) in a Xilinx Virtex-5 FPGA [3]:

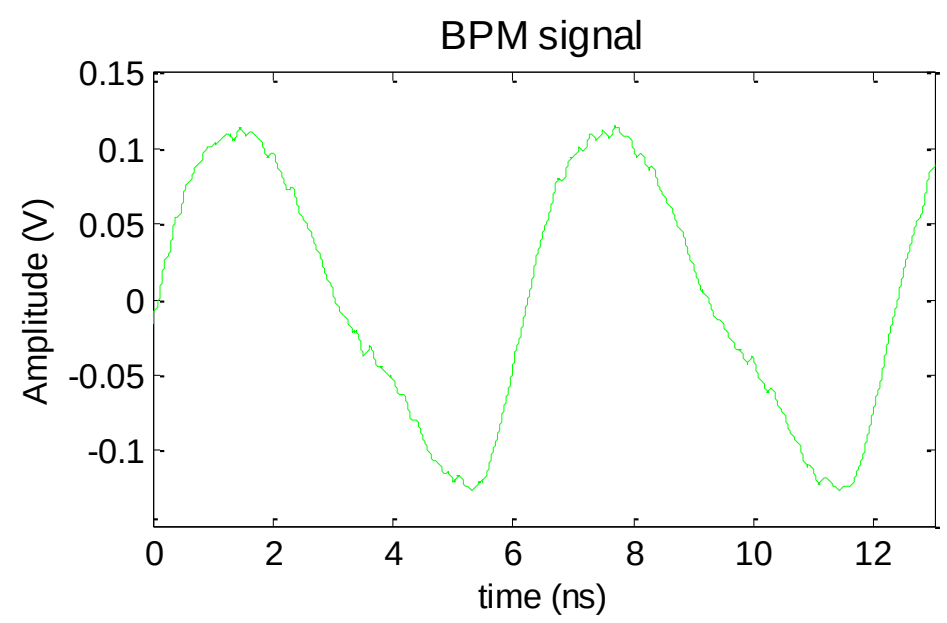

Fig. 1. The waveform of BPM output signal

$$Phase = \arctan\left( I/Q \right) \quad (1)$$

$$Amplitude = \sqrt{\left( I^2 + Q^2 \right)} \quad (2)$$

With the amplitude information, the beam position can be further obtained based on the Δ/Σ principle [19], as in (3) and (4):

$$X = K_X (V_R - V_L)/(V_R + V_L) - X_{offset} \quad (3)$$

$$Y = K_Y (V_T - V_B)/(V_T + V_B) - Y_{offset} \quad (4)$$

where $V_R$, $V_L$, $V_T$ and $V_B$ refer to the signal amplitudes from the four BPM pickups which are placed in the directions of right, left, top and bottom, as in Fig. 2 marked as A, B, C and D.

Manuscript received May 27, 2016. This work was supported in part by the National Natural Science Foundation of China under Grant 11205153 and 10875119, in part by the Knowledge Innovation Program of the Chinese Academy of Sciences under Grant KJCX2-YW-N27, in part by the Fundamental Research Funds for the Central Universities under Grant WK2030040029, and in part by the CAS Center for Excellence in Particle Physics (CCEPP).

The authors are with the State Key Laboratory of Particle Detection and Electronics, University of Science and Technology of China, Hefei, 230026; and Modern Physics Department, University of Science and Technology of China, Hefei, 230026, China (telephone: 086-0551-63607746, corresponding author: Lei Zhao, e-mail: zlei@ustc.edu.cn).

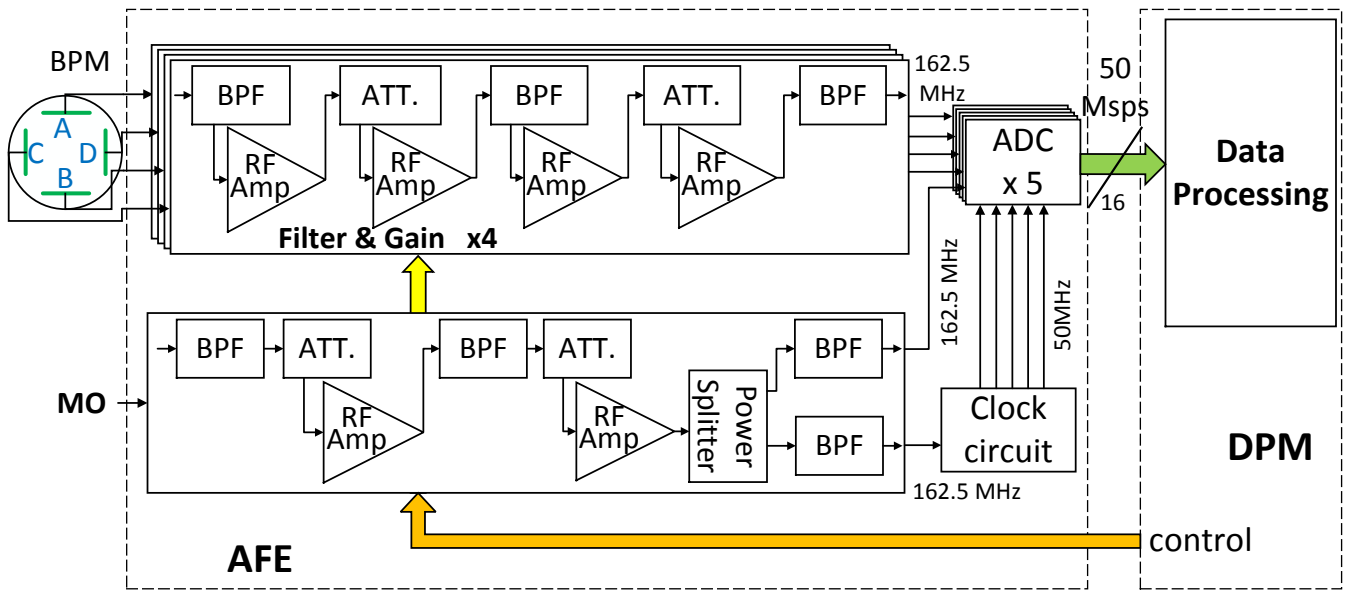

Fig. 2. Block diagram of the BPPM system.

Since the beam signals are converted to digital data, the calculation process can be implemented with the algorithms integrated in one FPGA device. However, channel mismatch and fluctuation of the gain and delay of analog circuits before A/D conversion are inevitable, which could cause measurement errors. Therefore, system calibration and correction are very important.

## II. Review of Beam Phase Correction Methods

As for the beam position measurement, beam position error is related with amplitude errors of four channels [5], so position correction can be easily conducted with calibration of the gain error of each channel. However, it is not an easy job for phase correction, and the correction methods vary with different detector types and phase calculation principles.

For the position-independent phase detector with a single pick-up such as Fast Current Transformers (FCT), the key task is to obtain the phase difference between two FCTs [6] [7] [9] [13]. So we can focus on the phase error (i.e. circuit delay) of two electronics channels that process these two FCT output signals, respectively. Large dynamic range is usually required for beam measurement systems, and it is inevitable that circuit delay changes with different gain settings. One simple method is to calibrate the delay variation for each channel, and then correct the error with Look-Up Tables (LUTs) [8] [15]. For some applications aiming for higher precision, two-dimensional LUTs can be used for phase error correction considering that the gain variation of one channel would impact on the delay of its adjacent channel [18]. As for the position-dependent phase detectors with multiple pick-ups, such as BPMs [10] [11], the situation is more complex. Take the BPM for example: the signals from the four pick-ups are required to be summed before phase calculation, in order to eliminate the detector phase dependence on position [11] [12]. As mentioned above, since the delay of each channel varies with different gain settings, there should be a complex procedure for phase correction. Besides, because the traditional method for signal summation is based on analog circuits [14] [16], the phase error of each electronics channel will be combined together in the output signal (i.e. the summed signal). For example, in the Bergoz BPM-AFE the "sum phase" error is around 3 degrees [14]. Under this situation, it is hard to implement phase correction. We have found no articles published to discuss this issue in detail.

In our system, the beam signals are directly A/D converted to digital IF signals, which makes it possible to correct the circuit phase error before summation. In this paper, we propose to combine the methods of signal transformation based on vector rotation and digital summation to achieve real-time phase correction with a correction precision better than the required ±0.5°. Analysis and estimation on the precision of this phase correction method are also presented in this paper.

## III. Correction Method

As shown in Fig. 2, signals from the four pick-ups of BPM are amplified and converted to sinusoid signals of 162.5 MHz through BPFs [3], and then directly digitized. As shown in Fig. 3, a "summed signal" is obtained through the logic in the FPGA, and then the phase information can be calculated. The final phase measurement result is just the difference between the phases of the summed signal and the reference signal (marked as "MO" in Fig. 2 & Fig. 3).

The sources which cause measurement errors are also illustrated in Fig. 3, which include gain and delay error.

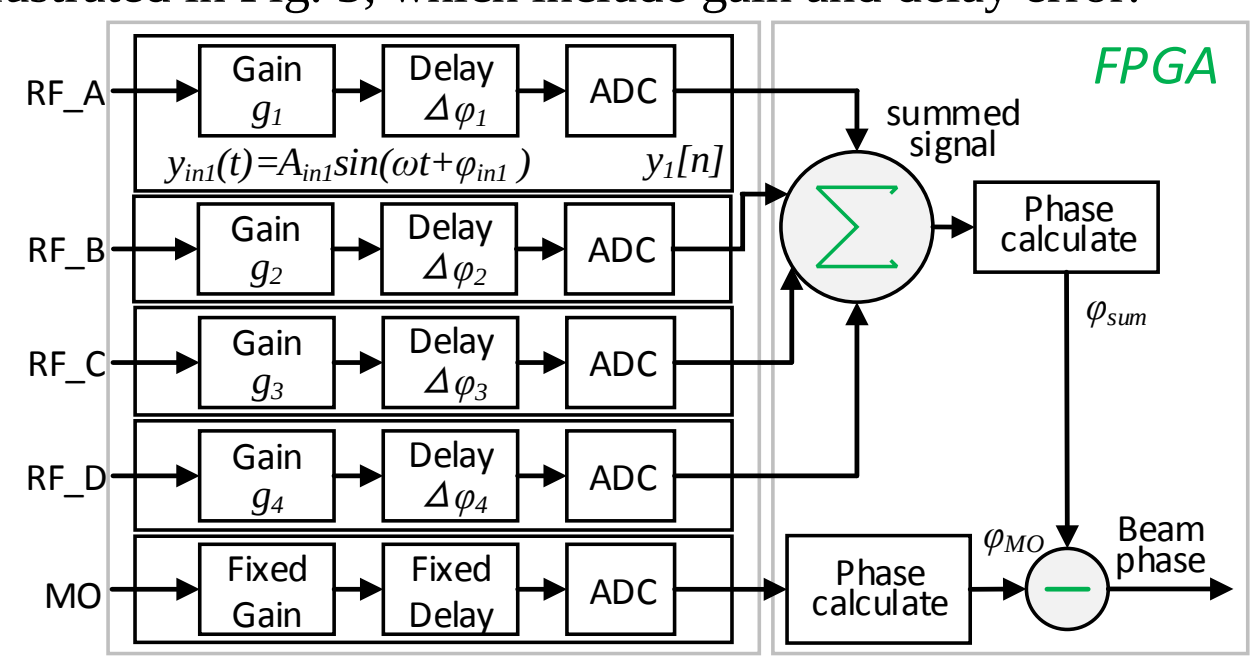

Fig. 3. System model with errors contributed by the circuits.

As shown in Fig. 3, we use $\varphi_{ink}$ and $A_{ink}$ to describe the input signal phase and amplitude, e.g. the input signal of channel 1 can be expressed as

$$y_{in1}(t) = A_{in1}\cdot\sin(\omega t+\varphi_{in1}) \qquad (5).$$

As mentioned above, the input signal is amplified to a constant amplitude (marked as "$A_{CON}$" in the following discussion) of around the FSR of ADC. It means that there exists a deterministic relationship between the circuit gain (marked as "$g_k$") and the input signal amplitude $A_{ink}$, i.e.

$$g_k\, A_{ink} = A_{CON} \qquad (6).$$

The output digital signals from the four ADCs can now be expressed as

$$\begin{aligned} y_k[n] &= g_k A_{ink}\sin(2\pi f_i \bullet t_n + \varphi_{ink} + \Delta\varphi_k) = A_{CON}\cdot\sin(2\pi n f_i / f_s + \varphi_k) \\ &= A_{CON}\cdot\sin(\frac{13\pi}{2}n+\varphi_k) = A_{CON}\cdot\sin(\frac{\pi}{2}n+\varphi_k) \quad (k=1,2,3,4) \end{aligned} \qquad (7),$$

where $f_i$ and $f_s$ are the frequencies of input signal (162.5 MHz) and sampling clock signal (50 MHz), $t_n$ is sampling time point, $\varphi_k$ is the output signal phase of the $k^{th}$ channel ($k$ represents the channel number), and $\Delta\varphi_k$ refers to the circuit delay. Of course, the sampling clock skews among different ADCs also contribute to $\Delta\varphi_k$, but they are constant errors and can be easily corrected. The phase of the "summed signal" ($\varphi_{sum}$) contains two parts – the real value ($\varphi_{insum}$) and the phase error ($\Delta\varphi_{sum}$) caused by the circuits in Fig. 3.

A straightforward idea to obtain the correct value of $\varphi_{insum}$ is calculating $\varphi_{sum}$, and then calibrating $\Delta\varphi_{sum}$ and correcting the phase measurement result by subtraction ($\varphi_{sum}$ - $\Delta\varphi_{sum}$). However, $\Delta\varphi_{sum}$ is the function of $A_{ink}$, $\varphi_k$ as in

$$\Delta\varphi_{sum} = \varphi_{sum} - \varphi_{insum} = f\left(A_{in1}, A_{in2}, A_{in3}, A_{in4}, \varphi_1, \varphi_2, \varphi_3, \varphi_4\right) \quad (8),$$

which means the calibration process would be quite complex. Considering the relationship in (6), (8) can also be expressed as a function of circuit gain ($g_k$) and $\varphi_k$. In the following discussion, we state only with the circuit gain settings.

$$\Delta\varphi_{sum} = f^*\left(g_1, g_2, g_3, g_4, \varphi_1, \varphi_2, \varphi_3, \varphi_4\right) \quad (9)$$

This is actually a feasible and traditional method in the situation of using one single pick-up, such as in [8] and [18]. However, in the situation of four pick-ups (e.g. BPM), the complexity of LUTs for correction will be greatly increased. According to (9), the LUT will be an 8-dimensional table, which consumes tremendous logic resources. For example, if this LUT covers a 60 dB dynamic range with a 1 dB step size, it requires a memory with a $60^8$ depth.

In this paper, we present a new correction method which features much lower resource consumption. The basic idea is to tune (i.e. correct) the signal waveform of each channel before summation, and implement real-time digital correction based on FPGA devices.

### A. *Phase correction method*

Since we intend to correct the signals of the four BPM channels, the core task is to compensate the gain error ($\Delta g_k$) and the delay difference ($\Delta\varphi_k$) of each channel. For the gain error, it is easy to correct through simple multiplication. As for the phase error, signal phase can be adjusted through digital delay filters, however, which is accompanied with high complexity. To simplify the correction process, we propose a new method in this paper, which is based on combination of RF signal IQ under-sampling technique and vector rotation.

As mentioned above, orthogonal I and Q arrays are obtained through IQ under-sampling process, and this makes it possible to implement phase correction by tuning the signal waveform through rotation. The whole phase correction consists of two steps: phase shift and amplitude transformation. If we consider the I and Q data as coordinates in the complex plane, phase shift is actually a rotation around the origin. And there is a 2x2 matrix $R_\theta$ that fully describes this rotation.

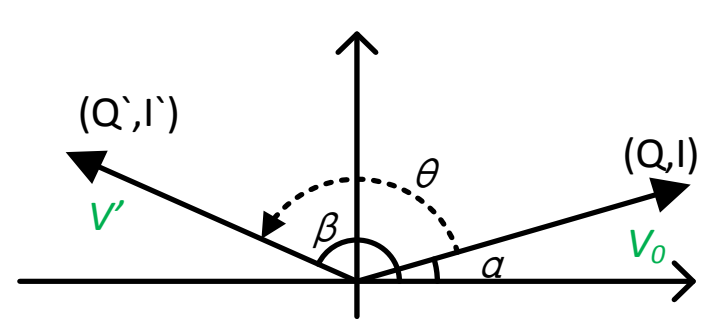

Fig. 4. Diagram of phase correction based on vector rotation.

This rotation is illustrated in Fig. 4. A given vector $\boldsymbol{V_0}$ is rotated by a counterclockwise angle $\boldsymbol{\theta}$ in a fixed coordinate system and the result vector $\boldsymbol{V'}$ can be expressed as

$$V' = \begin{pmatrix} Q' \\ I' \end{pmatrix} = R_\theta \cdot V_0 = \begin{pmatrix} \cos\theta & -\sin\theta \\ \sin\theta & \cos\theta \end{pmatrix} \begin{pmatrix} Q \\ I \end{pmatrix} \quad (10).$$

To simplify implementation of the above process on FPGA devices, we employed the COordinate Rotational DIgital Computer (CORDIC) IP core [17].

To guarantee a good phase correction precision, there is another issue that deserves attention. As mentioned above, the input signals are amplified to a constant amplitude ($A_{CON}$) before A/D conversion, as in (6). Therefore, in the calibration process, we can test the delay variations of each channel with different gains (while the input amplitudes are set to the corresponding values of $A_{CON}/g_k$) and establish LUTs; then in the correction process, we rotate the signal back according to the gain setting as the index of the LUTs. To simplify the gain controlling process, it is required to set the gains of four channels to an identical value, based on the principle to make the maximum amplitude among the four amplified signals approximate the FSR of ADC. For example, if the input amplitude of the 2nd channel is maximum among ($A_{in1}$, $A_{in2}$, $A_{in3}$, $A_{in4}$), the gains of all the four channels would be set to $g_1 = g_3 = g_4 = g_2 = A_{CON}/A_{in2}$. In this case, as for the 1st, 3rd and 4th channel, the input amplitudes would be smaller than $A_{CON}/g_k$ ($k = 1, 3, 4$). Considering that the circuit delay is influenced not only by gain setting but would also vary when the input signal amplitude changes, additional phase error would probably occur. However, since we use a summed signal to calculate the final phase result, these smaller signals contribute to phase calculation result with a lower weighting factor. So this error would be limited to a small range.

To simplify the analysis, we consider the summation of two channels. The output signals of these two channels are defined as vectors $\boldsymbol{A}$ and $\boldsymbol{B}$. And their amplitudes are marked as "*Amp_A"* and "*Amp_B"*, respectively, as shown in Fig. 5.

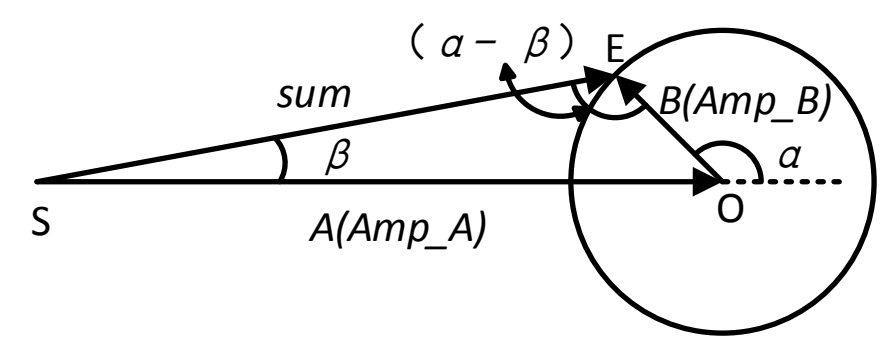

Fig. 5. Diagram of summation of two vectors.

The relationship between the phases of the summed signal (marked as "$\beta$") and the vector $\boldsymbol{B}$ (marked as "$\alpha$") can expressed as in

$$\frac{\sin\beta}{Amp_B} = \frac{\sin(\alpha-\beta)}{Amp_A} \quad (11)$$

If phase error (i.e. $\Delta\alpha$) exists in vector $B$, the phase error of the summed signal (i.e. $\Delta\beta$) can be derived, as in

$$\frac{\Delta\beta}{\Delta\alpha} = \frac{\dfrac{Amp_B}{Amp_A} + \cos\alpha}{\dfrac{Amp_A}{Amp_B} + \dfrac{Amp_B}{Amp_A} + 2\cos\alpha} \quad (12)$$

To estimate this error in the BPPM electronics, we tested the phase error of a typical channel. We fixed the gain of the channel and changed the input signal amplitude from -60 dBm to 0 dBm. Then we observed the variation of the phase measurement result, which is just the phase error, i.e. $\Delta\alpha$ in (12). The result is shown in Fig. 6.

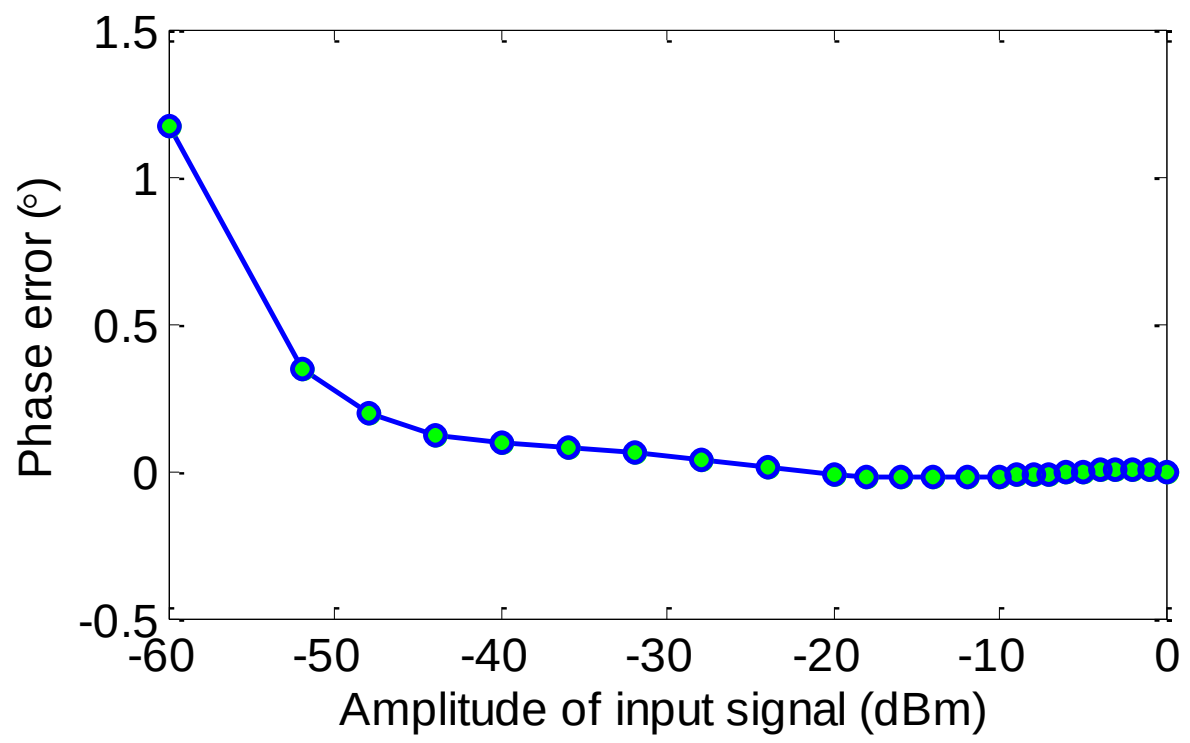

Fig. 6. Phase error of a typical channel with different input amplitudes under a constant gain setting.

Combining (12) with Fig. 6, the final phase error of the summed signal can be estimated, as in Fig. 7, indicating that a good phase precision can be obtained.

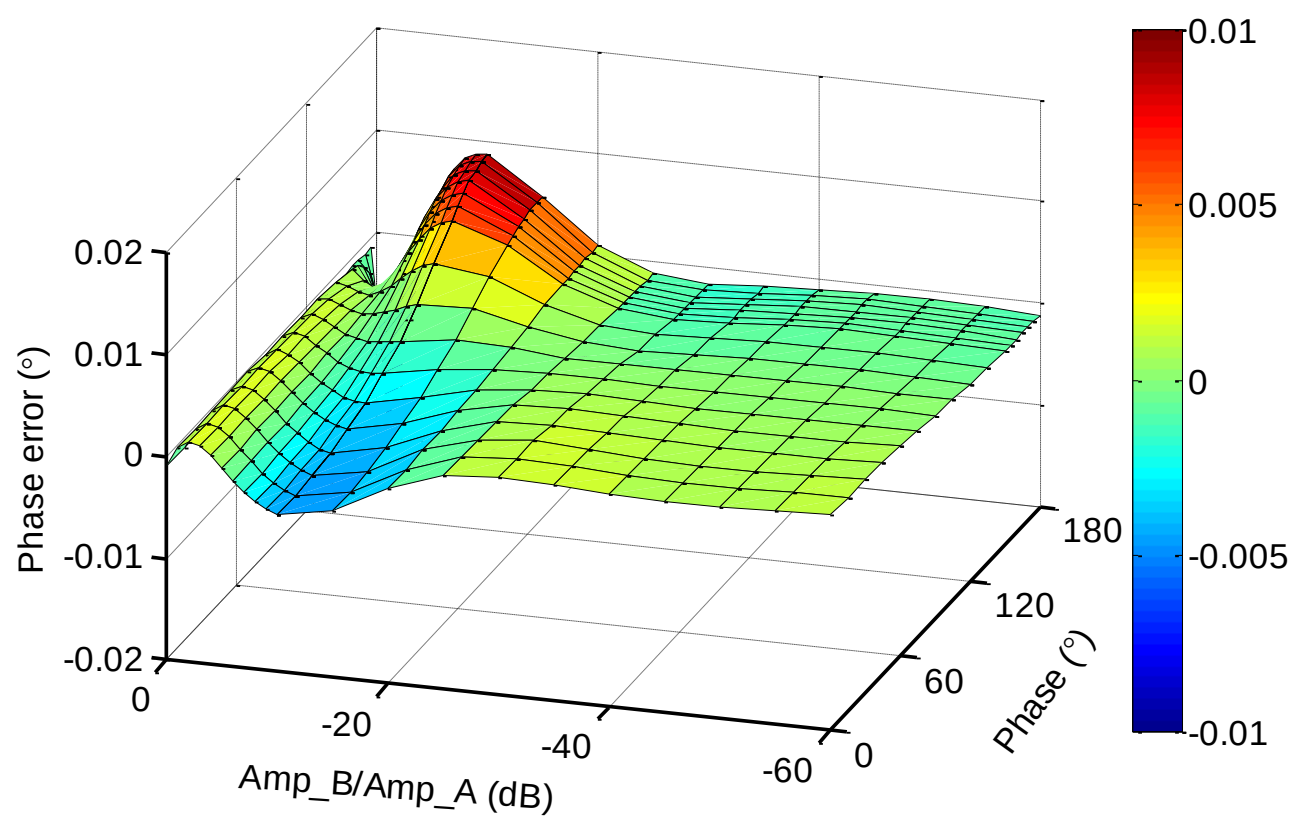

Fig. 7. Phase error estimation of the summed signal.

The above analysis demonstrates that the choice of employing a uniform gain value for all four channels will result in good phase precision with our correction method applied.

## B. *Logic design*

We implemented real-time correction algorithms in FPGA based on the above principle. The block diagram of the logic design is shown in Fig. 8. The whole phase correction algorithms consist of two parts -- phase correction and amplitude correction blocks. The input signals of the four channels, marked as "*IQ ChA*, *ChB*, *ChC* and *ChD*" are processed for phase correction through IQ rotation based on CORDIC algorithms. Since the delay (i.e. phase value) varies with different gains of the channels, we use only a one-dimensional LUT to record them. The second step is amplitude correction, and another one-dimensional LUT will be used to record the gain errors with different gain settings. This correction process is quite simple, which is implemented by using multipliers. The latency of the above phase correction algorithm is 500 ns. After these two steps, the signals of all four channels are corrected, which are ready for digital summation.

The corrected signals of four channels are then summed together, and then we can calculate the final phase measurement result. Meanwhile, the four corrected signals are also used to calculate the amplitudes of four channels, then the beam position can also be obtained base on Δ/∑ principle.

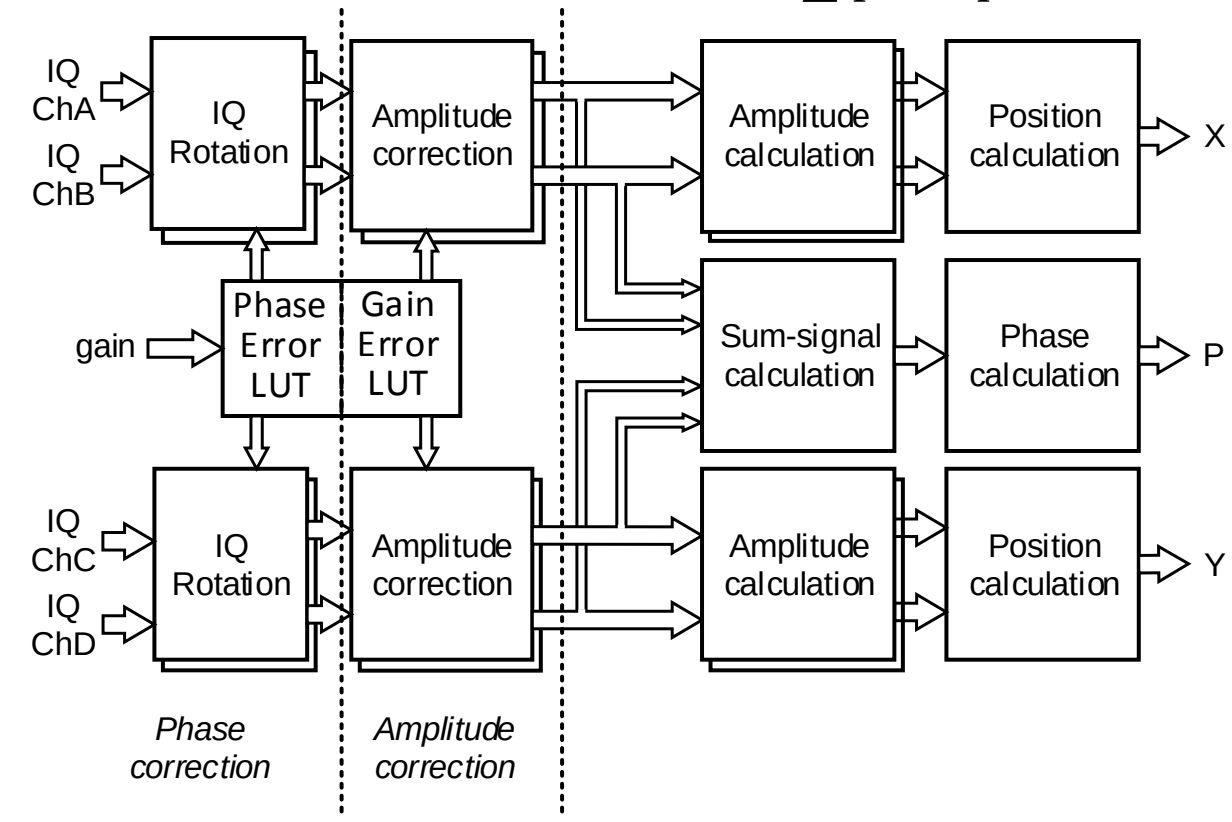

Fig. 8. Block diagram of the position and phase calculation logic.

The two one-dimensional LUTs only consume a memory of 9216 bits (half a Block Ram). Both phase and gain error values are 18 bits wide respectively, and they are combined to 36-bit data, which just fits the data width of the Block Ram. The IQ rotation, phase and amplitude calculation are all based on the CORDIC algorithms, so the logic structure is greatly simplified. The source consumption of the overall correction algorithm in Virtex-5 XC5VLX155T FPGA is listed in Table 1.

Table 1
Resource Utilization

| Resource | Used | Available | Utilization |
|---|---|---|---|
| Slice Registers | 7146 | 97280 | 7.4% |
| Slice LUTs | 6660 | 97280 | 6.9% |
| Block Ram | 1 | 218 | 0.4% |
| DSP48Es | 12 | 128 | 9.4% |

# IV. Test Results

To evaluate the performance of the correction method, we conducted tests in the laboratory. In the first step, we calibrated the phase and gain errors of the four channels. As shown in Fig. 9, we used a high quality RF signal source ROHDE&SCHWARZ SMA100A to generate a 162.5 MHz sinusoid signal. The output signal was then split into four paths as the RF input signals of the BPPM. The signal source also generated a synchronized MO signal for the BPPM. Then we tested phase values (delay) of four channels referenced to the MO signal with different gain settings. We also employed a Vector Network Analyzer Agilent E5071C to conduct the same tests with the results as reference, and then the phase and gain errors were calibrated. Fig. 10 shows the system under test.

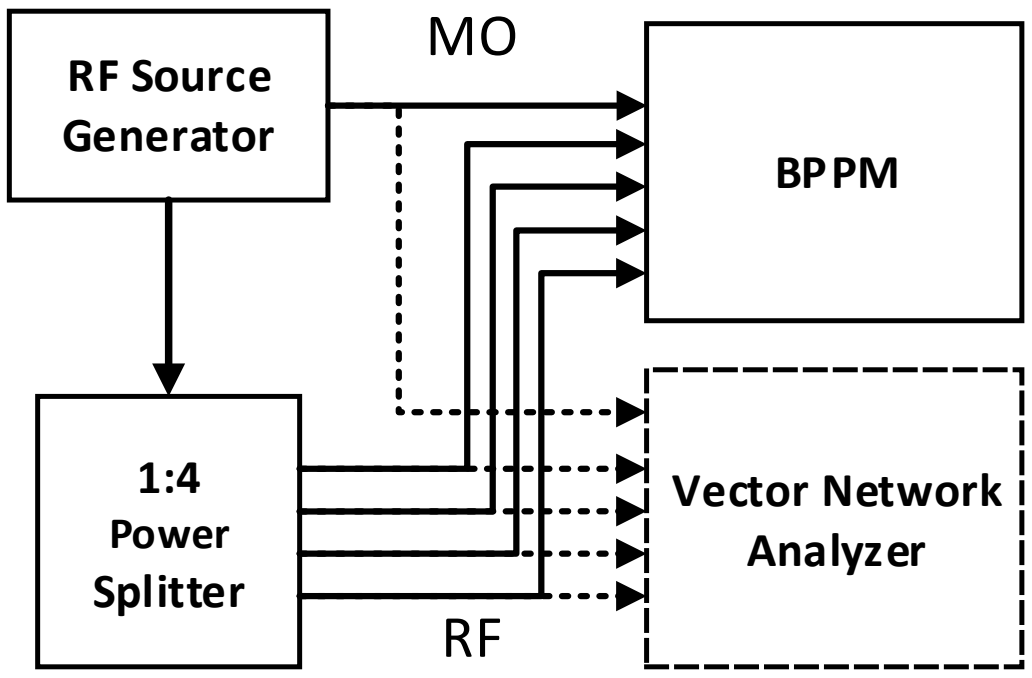

Fig. 9. System setup for phase calibration.

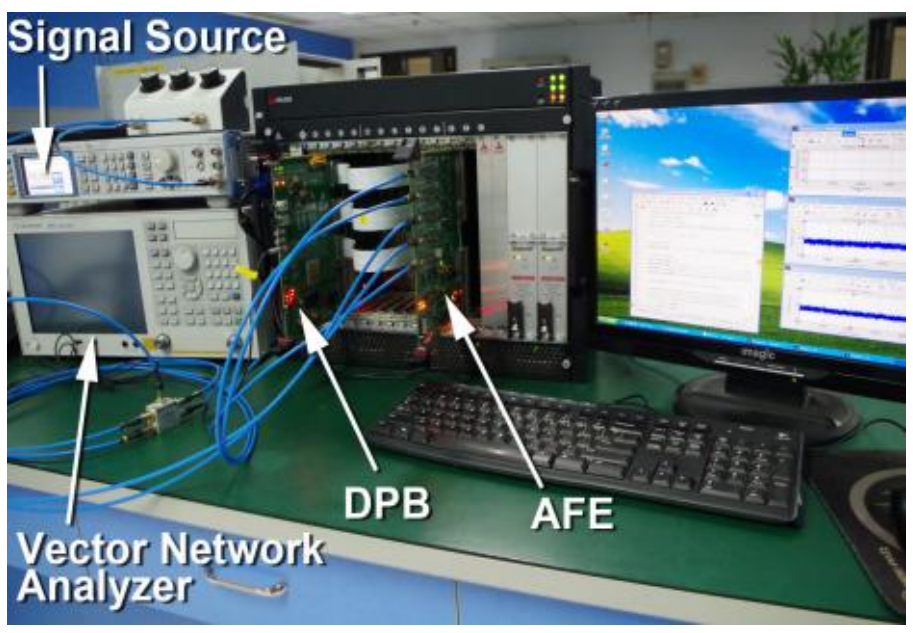

Fig. 10. System under test.

The phase calibration results are shown in Fig. 11. The curve discontinuities in Fig. 11 are caused by the inconsistency of attenuators in the BPPM system when their digital controlling bits transit over special codes. Using the results in Fig. 11, we can now establish the correction LUT in Fig. 8. Then we conducted tests to evaluate the effect of our correction logic. In the tests, we fixed the amplitude of one channel (Channel A), and swept the amplitudes of the other channels (Channel B, C, D) in the range of 60 dB. This is actually a more severe situation compared to the real application.

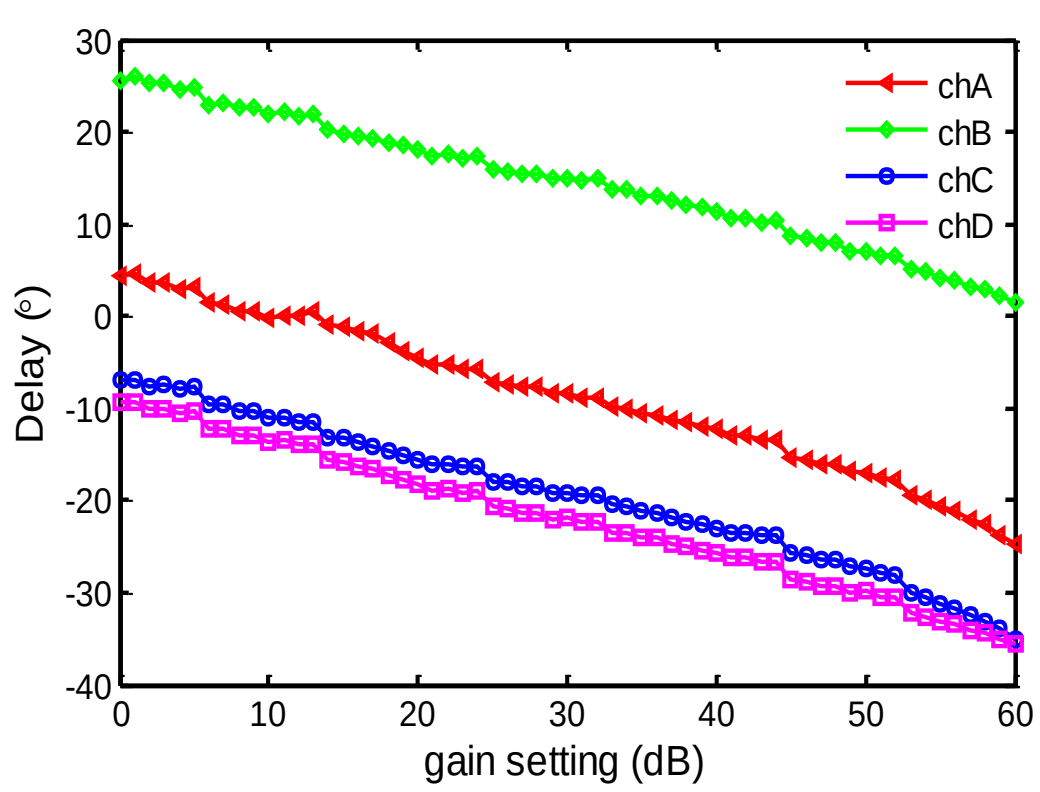

Fig. 11. Phase calibration results of four channels with different gain settings

Fig. 12 shows the test results over an amplitude range from -60 dBm to 0 dBm (the amplitudes of Channel B, C, D), in which 64k samples are averaged to obtain final phase value. The results in Fig. 12 indicate the phase error was reduced to ±0.3° with the correction applied, beyond application requirement.

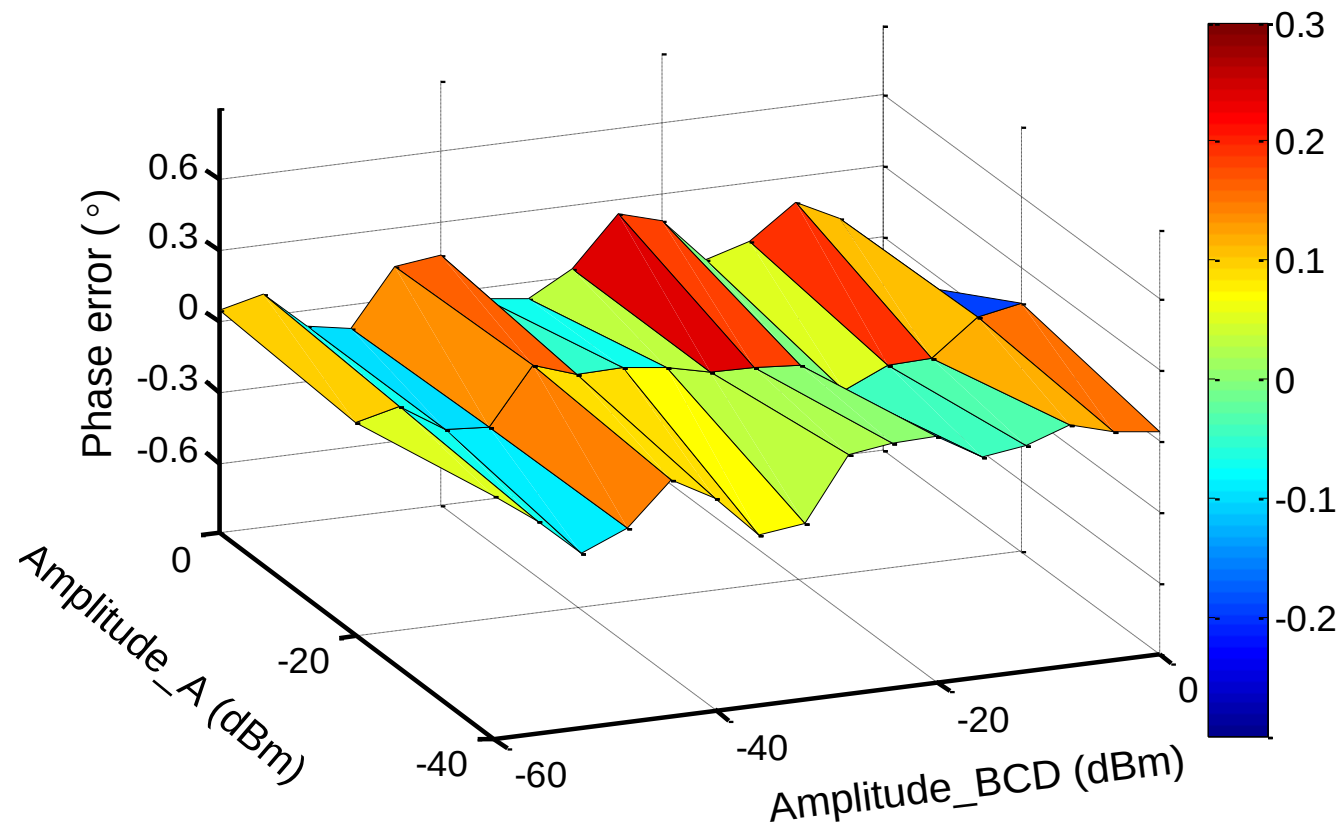

Fig. 12. The phase error test results after correction.

## V. Conclusion

Based on fully digital hardware design and digital waveform tuning through rotation and summation, real-time phase error correction is achieved within one single FPGA. The principle of this method is studied and analyzed. And its performance is evaluated through tests in the laboratory. Test results indicate that a phase correction precision is better than ±0.3° in the dynamic range from -60 dBm to 0 dBm, which is beyond application requirement. This fully digital beam phase correction method has an advantage of real-time signal processing and reducing system complexity. It provides a solution that could be applied to BPPMs, which are widely used in the accelerator beam diagnostics.